\documentclass[twocolumn]{aastex631}

\newcommand\ml{$M/L$}

\newcommand\lgm{$\log M/M_\odot$}

\shorttitle{Stellar masses and \ml~of $z>7$ galaxies from GLASS-JWST-ERS}
\shortauthors{Santini et al.}
\graphicspath{{./}{figs/}}

\begin{document}


\title{Early results from GLASS-JWST. XI: Stellar masses and mass-to-light ratio of $z>7$ galaxies  \footnote{Released on ...}}

\correspondingauthor{Paola Santini}
\email{paola.santini@inaf.it}

\author[0000-0002-9334-8705]{P. Santini}
\affiliation{INAF - Osservatorio Astronomico di Roma, via di Frascati 33, 00078 Monte Porzio Catone, Italy}

\author[0000-0003-3820-2823]{A. Fontana}
\affiliation{INAF - Osservatorio Astronomico di Roma, via di Frascati 33, 00078 Monte Porzio Catone, Italy}

\author[0000-0001-9875-8263]{M. Castellano}
\affiliation{INAF - Osservatorio Astronomico di Roma, via di Frascati 33, 00078 Monte Porzio Catone, Italy}

\author[0000-0003-4570-3159]{N. Leethochawalit}
\affiliation{School of Physics, University of Melbourne, Parkville 3010, VIC, Australia}
\affiliation{ARC Centre of Excellence for All Sky Astrophysics in 3 Dimensions (ASTRO 3D), Australia}
\affiliation{National Astronomical Research Institute of Thailand (NARIT), Mae Rim, Chiang Mai, 50180, Thailand}

\author[0000-0001-9391-305X]{M. Trenti}
\affiliation{School of Physics, University of Melbourne, Parkville 3010, VIC, Australia}
\affiliation{ARC Centre of Excellence for All Sky Astrophysics in 3 Dimensions (ASTRO 3D), Australia}

\author[0000-0002-8460-0390]{T. Treu}
\affiliation{Department of Physics and Astronomy, University of California, Los Angeles, 430 Portola Plaza, Los Angeles, CA 90095, USA}

\author{D. Belfiori}
\affiliation{INAF - Osservatorio Astronomico di Roma, via di Frascati 33, 00078 Monte Porzio Catone, Italy}

\author{S. Birrer}
\affiliation{Kavli Institute for Particle Astrophysics and Cosmology and Department of Physics, Stanford University, Stanford, CA 94305, USA}
\affiliation{SLAC National Accelerator Laboratory, Menlo Park, CA, 94025}

\author{A. Bonchi}
\affiliation{Space Science Data Center, Italian Space Agency, via del Politecnico, 00133, Roma, Italy}

\author[0000-0001-6870-8900]{E. Merlin}
\affiliation{INAF - Osservatorio Astronomico di Roma, via di Frascati 33, 00078 Monte Porzio Catone, Italy}

\author[0000-0002-3407-1785]{C. Mason}
\affiliation{Cosmic Dawn Center (DAWN)}
\affiliation{Niels Bohr Institute, University of Copenhagen, Jagtvej 128, 2200 København N, Denmark}

\author[0000-0002-8512-1404]{T. Morishita}
\affiliation{IPAC, California Institute of Technology, MC 314-6, 1200 E. California Boulevard, Pasadena, CA 91125, USA}

\author[0000-0001-6342-9662]{M. Nonino}
\affiliation{INAF - Osservatorio Astronomico di Trieste, Via Tiepolo 11, I-34131 Trieste, Italy}

\author[0000-0002-7409-8114]{D. Paris}
\affiliation{INAF - Osservatorio Astronomico di Roma, via di Frascati 33, 00078 Monte Porzio Catone, Italy}

\author[0000-0003-4067-9196]{G. Polenta}
\affiliation{Space Science Data Center, Italian Space Agency, via del Politecnico, 00133, Roma, Italy}

\author[0000-0002-6813-0632]{P. Rosati}
\affiliation{Dipartimento di Fisica e Scienze della Terra, Università degli Studi di Ferrara, Via Saragat 1, I-44122 Ferrara, Italy}
\affiliation{INAF - OAS, Osservatorio di Astrofisica e Scienza dello Spazio di Bologna, via Gobetti 93/3, I-40129 Bologna, Italy}

\author[0000-0002-8434-880X]{L. Yang}
\affiliation{Kavli Institute for the Physics and Mathematics of the Universe, The University of Tokyo, Kashiwa, Japan 277-8583}

\author[0000-0001-5984-0395]{M. Bradac}
\affiliation{
University of Ljubljana, Department of Mathematics and Physics, Jadranska ulica 19, SI-1000 Ljubljana, Slovenia}
\affiliation{
Department of Physics and Astronomy, University of California Davis, 1 Shields Avenue, Davis, CA 95616, USA}

\author[0000-0003-2536-1614]{A. Calabr\`o}
\affiliation{INAF - Osservatorio Astronomico di Roma, via di Frascati 33, 00078 Monte Porzio Catone, Italy}

\author[0000-0002-6317-0037]{A.~Dressler}
\affiliation{The Observatories, The Carnegie Institution for Science, 813 Santa Barbara St., Pasadena, CA 91101, USA}

\author[0000-0002-3254-9044]{K. Glazebrook}
\affiliation{Centre for Astrophysics and Supercomputing, Swinburne University of Technology, PO Box 218, Hawthorn, VIC 3122, Australia}

\author[0000-0001-9002-3502]{D. Marchesini}
\affiliation{Department of Physics and Astronomy, Tufts University, 574 Boston Ave., Medford, MA 02155, USA}

\author[0000-0002-9572-7813]{S. Mascia}
\affiliation{INAF - Osservatorio Astronomico di Roma, via di Frascati 33, 00078 Monte Porzio Catone, Italy}

\author[0000-0003-2804-0648 ]{T. Nanayakkara}
\affiliation{Centre for Astrophysics and Supercomputing, Swinburne University of Technology, PO Box 218, Hawthorn, VIC 3122, Australia}

\author[0000-0001-8940-6768 ]{L. Pentericci}
\affiliation{INAF - Osservatorio Astronomico di Roma, via di Frascati 33, 00078 Monte Porzio Catone, Italy}

\author[0000-0002-4140-1367]{G. Roberts-Borsani}
\affiliation{Department of Physics and Astronomy, University of California, Los Angeles, 430 Portola Plaza, Los Angeles, CA 90095, USA}

\author[0000-0002-9136-8876]{C.~Scarlata}\affiliation{
School of Physics and Astronomy, University of Minnesota, Minneapolis, MN, 55455, USA}

\author[0000-0003-0980-1499]{B. Vulcani}
\affiliation{INAF Osservatorio Astronomico di Padova, vicolo dell'Osservatorio 5, 35122 Padova, Italy}

\author[0000-0002-9373-3865]{Xin Wang}
\affil{Infrared Processing and Analysis Center, Caltech, 1200 E. California Blvd., Pasadena, CA 91125, USA}



\begin{abstract}
We exploit James Webb Space Telescope ({\em JWST}) NIRCam observations from the GLASS-JWST-Early Release Science  program  to investigate galaxy stellar masses 
at $z>7$. We first show that {\em JWST} observations reduce the uncertainties on the stellar mass by a factor of at least 5-10, when compared with the highest quality data sets available to date. We then study the UV mass-to-light ratio, finding that galaxies exhibit a two orders of magnitude  
range of $M/L_{UV}$ values for a given 
luminosity, indicative of  a broad variety of physical conditions and star formation histories. As a consequence, previous estimates of the cosmic star stellar mass density -- based on an average correlation between UV luminosity and stellar mass -- can be biased by as much as a factor of $\sim$6.  
Our first exploration demonstrates that  {\em JWST} represents a new era in our understanding of  stellar masses at $z>7$, and therefore of the growth of galaxies prior to cosmic reionization. 
\end{abstract}

\
\keywords{galaxies: evolution --- galaxies:
  high-redshift --- galaxies: fundamental parameters}


\section{Introduction}\label{sec:intro}

Stellar mass is one of the 
most fundamental physical properties describing galaxies. Reliable measurements of this galaxy property are crucial to understand the overall picture of galaxy formation and evolution. 
Stellar masses of galaxies are usually calculated by fitting stellar population synthesis models to  broad band multiwavelength photometry. 
The accuracy of these  estimates hinges on the availability of  rest-frame optical and near-infrared (NIR) photometry  \citep[e.g.][]{santini12a,paulino-afonso22}, which trace the bulk of the emission from stars that most contribute to the stellar mass. This wavelength range is exquisitely sampled by HST up to $z\sim 3$. At higher redshift Spitzer has been widely used but is able to provide information on the very brightest galaxies only, as it is limited by low sensitivity and poor angular resolution (compared to HST), and blending issues, especially in crowded fields. 

As a consequence of the observational limitations prior to {\em JWST}, most studies of  
galaxies at $z\gtrsim 6$ to date converted rest-frame UV fluxes into mass estimates through  
an average mass-to-light ratio ($M/L_{UV}$, for simplicity referred to as \ml~ in the following), rather than computing the stellar masses on a source-by-source base 
\citep[e.g.][]{gonzalez11,song16,kikuchihara20}.

Another limitation prior to {\em JWST} is that, without rest-frame optical and NIR observations one could miss a large fraction of intrinsically red galaxies, i.e. those galaxies that are faint in the UV, such as evolved or dust--obscured systems. Current high redshift samples are  therefore likely to be severely incomplete in terms of this population. As a consequence,  \ml~are  calibrated (almost only) on UV-detected, Lyman Break Galaxies (LBG), and likely inappropriately applied to a larger variety of galaxy populations \citep[as discussed in][]{grazian15}.

In this Letter we use the power of {\em JWST} to carry out the first determination of stellar masses at high redshift using rest-frame optical data. By comparing with previous work we demonstrate the importance of this new observing window. 
In practice, we exploit the first NIRCam \citep{rieke05} observations of the GLASS-ERS 1324 program \citep{treu22}, and measure stellar masses of  $z\gtrsim 7$ galaxies, analyze their \ml,  
and critically assess the adoption of the UV emission as a mass tracer. 

We emphasize that this is a first look at this important issue, based on the current understanding of the data quality and calibration of NIRCam data\footnote{The "pedigree" of the data products adopted here is   \texttt{CAL\_VER 1.6.0}; \texttt{CRDS\_CTX jwst\_0942.pmap} (released  on July 29 2022).}. 
We expect to significantly improve our data set and analysis in the near future. 

This Letter is organized as follows: 
we describe the dataset and methodology in Sect.~\ref{sec:data}, present our results in Sect.~\ref{sec:results}, and summarize our findings in Sect.~\ref{sec:summary}. 
We adopt  the standard $\Lambda$ Cold Dark Matter concordance cosmological model  ($H_0 = 70$ km s$^{-1}$ Mpc$^{-1}$, $\Omega_M = 0.3$, and $\Omega_\Lambda = 0.7$) and a \cite{chabrier03} 
(IMF). Magnitudes are given in the AB system.

\section{Data set and methods}\label{sec:data}

We use the first set of NIRCam observations taken on June 28-29 2022 as part of the GLASS-ERS 1324 program in 7 wide filters (F090W, F115W, F150W, F200W, F277W, F356W and F444W). These data were taken in parallel to NIRISS \citep{doyon12} observations \citep[Paper I]{roberts-borsani22}, targeting the Abell2744 cluster, with the  NIRCam fields centered at RA=0:14:02.6142, Dec=-30:21:38.793 and RA=0:13:58.4752, Dec=-30:18:52.094. They were analysed through the {\em JWST} pipeline\footnote{\url{https://jwst-docs.stsci.edu/jwst-science-calibration-pipeline-overview}}, slightly modified as described by \citet[Paper II]{merlin22}. 
Source detection was carried out on the F444W band, while aperture fluxes were computed in the other bands on the PSF-matched images. We use corrected aperture photometry, computed as explained by \cite{merlin22}. 

Galaxies were selected based on color and photometric redshift criteria, as described in two companion papers. The $7<z <9$ sample is described by \citet[Paper X]{leethochawalit22}, while \citet[Paper III]{castellano22} focuses on $z>9$. In total,  14 candidates were selected based on their colors and  signal-to-noise-ratio (SNR), and additional   5  were included based on the photo-$z$ selection.  
We refer to the two companion papers mentioned above for further details. Our total sample comprises   19  galaxies, 
spanning a range in redshfit from  6.9  to  12.1, with a  mean (median)  value of  $\sim$8.6 (8.1), over an effective area of $\sim$6.6 arcmin$^2$ (which is smaller than the NIRCam field of view due to non perfect overlap of the images in all bands). 

For one of the galaxies in our sample, namely GHZ2, the photometry has been calculated ad-hoc to avoid contamination from a bright close-by galaxy. The photometric measurement of this source is described in detail in \cite{castellano22}.

\subsection{Stellar mass estimates}

\begin{figure}[t!]
    \centering
\includegraphics[width=9cm,]{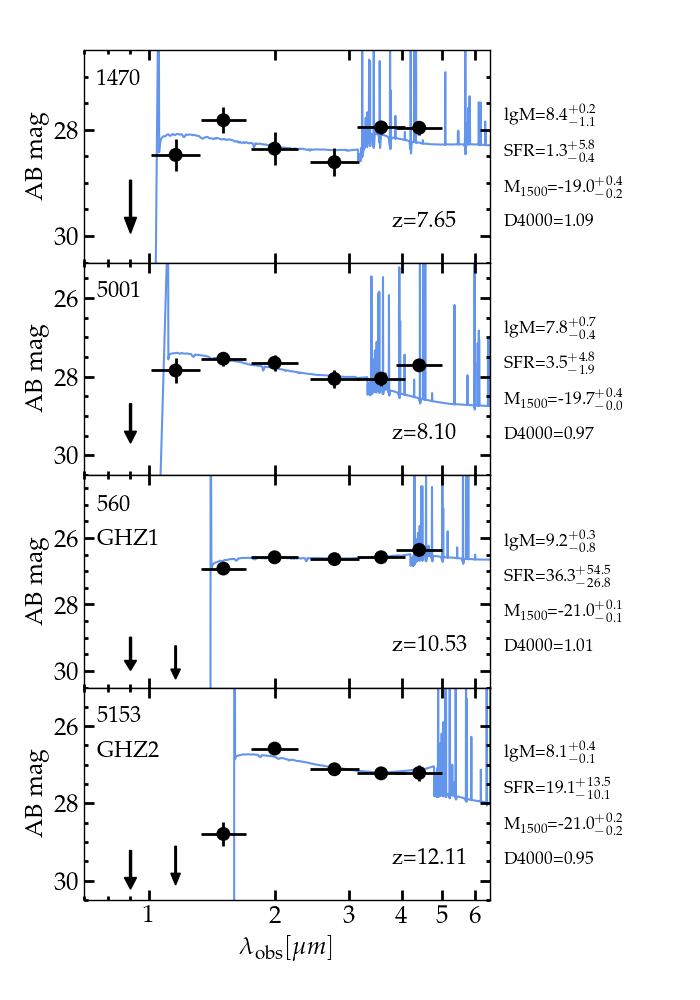}
\caption{Spectral Energy Distribution and best-fit template of four of our candidates (we note that Ly$\alpha$ is not accounted for in the fit).  Upper limits are shown at a 2$\sigma$ confidence level. The two top panels show $z\sim 7.5-8$ galaxies, exhibiting large ({\it top}) and low ({\it central top})  D4000 and \ml. The two bottom panels show the two most robust $z>10$ galaxies identified by \cite{castellano22}. 
}
\label{fig:seds}
\end{figure}

Stellar masses and the other physical parameters, including the observed absolute UV magnitudes at 1500\AA~($M_{1500}$), were measured 
by fitting synthetic stellar templates to the 7-band NIRCam photometry  with \textsc{zphot }\citep{fontana00}. When the nominal flux errors are smaller than 0.05 mags, we set a minimum photometric uncertainty in all bands corresponding to this threshold to account for uncertainties in the NIRCam calibration. 
We adopt in the following the photometric redshifts computed by \cite{leethochawalit22} and  \cite{castellano22}. We fixed the redshift to the \textsc{EAzY} $z_{peak}$ value,  with the exception of  two of the $z>9$ color selected galaxies (GHZ3 and GHZ5 in \citealt{castellano22}), for which \textsc{EAzY}  prefers a lower redshift solution while  \textsc{zphot}  fits them at $z>9$. We use the \textsc{zphot} redshifts for these sources. 
We built the stellar library following the assumptions of \cite{merlin21}. 
We adopt \cite{bc03} models, 
including nebular emission lines according to \cite{castellano14} and  \cite{schaerer09}. We do not include Ly$\alpha$ emission because it is most likely absorbed by the Intergalactic Medium at these redshifts, although for clarity it is drawn in the Spectral Energy Distributions (SED) shown in Fig.~\ref{fig:seds}. 
We assume delayed exponentially declining Star Formation Histories  (SFH($t$)$\propto (t^2/\tau) \cdot \exp(-t/\tau)$)  with $\tau$ ranging from 0.1 to 7 Gyr (the adoption of more complicated SFHs, e.g. including recent bursts, will be addressed in a future analysis).  We let the age range  from 10 Myr to the age of the Universe at each galaxy redshift. Metallicity is allowed to be 0.02, 0.2 and 1 times Solar, 
and dust extinction is assumed to follow a \cite{calzetti00} law with E(B-V) varying from 0 to 1.1. 
We compute $1\sigma$  uncertainties on the physical parameters by retaining for each object the minimum and maximum fitted masses among all the solutions with a probability $P(\chi^2)>32\%$ of being correct, both fixing the redshift to the best-fit value and allowing it to vary within its 1 and 2$\sigma$ ranges. For this analysis, for simplicity, we consider uncertainties at fixed redshifts. We will explore any residual degeneracy with redshift in future work.

We have  compared the stellar masses computed with \textsc{zphot} with those computed at the very same redshift with \textsc{Bagpipes} \citep{carnall18}   assuming a log-normal SFH, used by \cite{leethochawalit22}. We find good agreement between the two, with some discrepancy only for a few low mass galaxies. These discrepancies arise in objects with very small 4000\AA~break (D4000). For these objects, the photometry at long wavelengths depends critically on the assumptions about emission lines, which are different for the two codes.  This is clearly a topic that will deserve further attention as larger samples of spectra at $z>7$ become available to calibrate the models.

We show in Fig.~\ref{fig:seds} a few examples of the SED and best-fit templates of our candidates:  two $z\sim 7.5-8$ candidates, one with high \ml~and a pronounced D4000  (ID1470) and another one with low \ml~ and a  D4000 smaller than 1  (ID5001), 
and the two most robust candidates at $z>10$, namely GHZ1 and GHZ2 presented by \cite{castellano22}. We note that two out of these four galaxies exhibit  D4000$<$1, which is indicative of extremely young stellar populations and a spectrum dominated by nebular continuum.  While the optical rest-frame is nicely sampled  at $z<10$ -- and also marginally at $z\sim 10$ -- at higher redshift we are limited to the spectral region blueward of the break. For these galaxies longer wavelength observations with MIRI will be a valuable addition. 

Remarkably, GHZ2 shows a  characteristic U-shape pattern \citep{carnall22} (i.e. an extremely blue UV slope with the Balmer break seen in emission) that allows us to nicely constrain its SED. Despite its extremely high redshift, its intrinsic luminosity allows for an accurate physical characterization. 

Modest lensing magnification is expected to be present in the  parallel fields \citep{medezinski16,bergamini22}, so that  the stellar masses and UV luminosities should be taken as lower limits. 
In this initial set of papers we neglect this effect. The issue will be revisited after the completion of the campaign. 

In any case, the main results of the present Letter are independent of the lensing magnification. In particular,  lensing affects masses and luminosities in the same way, so the \ml~ratio is free from magnification uncertainties.
Furthermore, relative uncertainties on the stellar mass are also independent of magnification.

\section{Results and discussion} \label{sec:results}

\subsection{Accuracy of stellar mass estimates}

\begin{figure}[t!]
    \centering
\includegraphics[width=9cm]{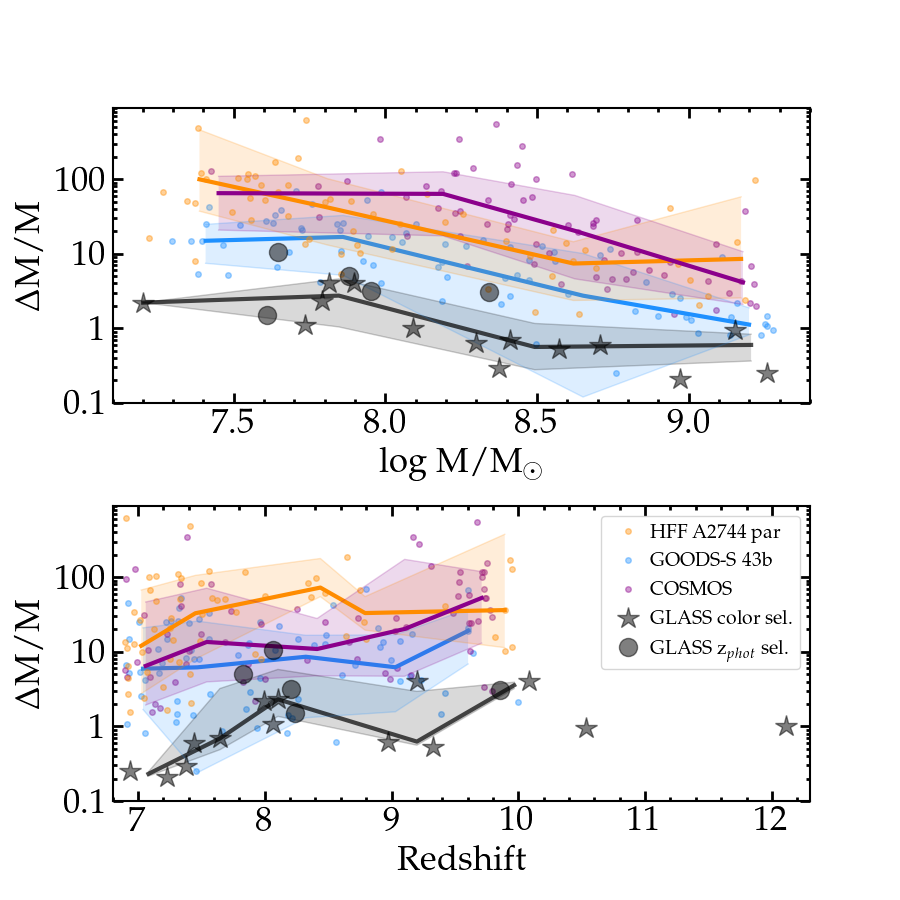}
\caption{Relative uncertainty on the stellar mass  as a function of stellar mass ({\it upper} panel) and redshift ({\it lower} panel). Large gray symbols show GLASS observations, with stars and circles indicating the color selected candidates  and the additional ones included via photo-$z$ criteria, respectively. Smaller blue, purple and orange 
symbols show the results for  CANDELS GOODS-S, CANDELS COSMOS 
and the parallel Hubble Frontier Field A2744, respectively, analysed in the same way as GLASS. For the comparison samples, we restrict to galaxies with $z>6.9$ and $7.2<\log M/M_\odot<9.3$. For all  four data sets, curves  
show medians values in bins of stellar mass and redshift, plotted at the median $x$-value. 
The binning, whose step is 0.75 in both stellar mass and redshift, is the same for the various data sets. Shaded region represent the area between the 16$^{th}$ and 84$^{th}$ percentiles of the distributions. 
}
              \label{fig:deltam}
\end{figure}

{\em JWST} observations allow for a significant reduction of the uncertainties on the stellar mass measurement. Figure~\ref{fig:deltam} shows the relative uncertainty $\Delta M/M$ at a 1$\sigma$ level as a function of stellar mass and redshift. The error $\Delta M = (M_{max}-M_{min})/2$ is computed by scanning all models with acceptable $\chi^2$ as described above.

The $\Delta M/M$ inferred from GLASS observations is compared with the uncertainty affecting the stellar masses in three of the highest quality  and most exploited data sets that were available before the advent of {\em JWST}. All of them are obtained with a combination of HST, Spitzer and ground--based data covering approximately the same spectral range of this work, but with much lower depth and SNR longward $1.6\mu$m. These are: 
{\it i)} the 43 band catalog of the CANDELS GOODS-S field by \cite{merlin21}, {\it ii)} the 43 band CANDELS COSMOS catalog of  \cite{nayyeri17} and {\it iii)} the 10 band catalog of  Hubble Frontier Field (HFF) A2744 parallel by \cite{merlin16} and \cite{castellano16}.  The 5$\sigma$ limiting magnitudes in the CH1 and CH2 IRAC bands are $25.5-25.6$ ({\it i}, total magnitudes), 
24.4 ({\it ii}, aperture photometry within a 1 FWHM radius) and $\sim 24.85$ ({\it iii}, calculated as described in \citealt{merlin16}), to be compared with the 29.3-29.7  depth (5$\sigma$ limiting point source magnitudes in 0.1\arcsec~radius) of the F356W and F444W GLASS observations \citep{merlin21}. 
In all cases, we restricted to the galaxies of similar redshift ($z>6.9$) and mass ($7.2<$ \lgm $<9.3$) as the GLASS sample.  In addition, we cleaned the catalogs by removing all sources flagged in the original catalogs due to photometric issues as well as spectroscopic and photometric stars. The latter were identified requiring SNR(H160)$>$10 and  either {\it a)} SExtractor \citep{bertin96} \textsc{class$\_$star}$>$0.95 or {\it b)} \textsc{class$\_$star}$>$0.8
and populating the stellar locus of the {\it BzK} diagram
\citep{daddi04}. 
For the GOODS-S and COSMOS catalogs, we also removed X-ray selected AGNs according to the official CANDELS selection.

We remark that the computation of the error on mass depends on the code and on the library adopted. For this reason, we have used exactly the same code and template library,  
but for safety the comparison should be considered as an estimate of the relative improvement that can be obtained via {\em JWST} optical rest-frame photometry.

Figure~\ref{fig:deltam} shows that, in the redshift and mass range considered here, {\em JWST} observations improve the accuracy on the stellar mass measurements by factors ranging from $\simeq 2$ to $\simeq$  30 
on average compared to measurements based on Spitzer, HST and ground based observations only, 
at a given stellar mass  or redshift. The improvement is more significant with respect to the HFF data set. The HFF catalog lacks IRAC CH3 and CH4 bands, which are instead included in the  CANDELS catalogs, and suffers from shorter exposure times with Spitzer. Therefore, the deeper HST photometry of  HFF compared to CANDELS is not complemented by similarly deeper IRAC CH1 and CH2 observations. In any case, we need to emphasize that the exposure times of the Spitzer data are significantly larger than those of {\em JWST} here, up to more than 100hr in the case of GOODS-S. 

Unsurprisingly, the accuracy improves for more massive objects, which are on average brighter and proportionally less affected by the (uncertain) contribution of nebular emission lines. 
The relative improvement compared to HFF is even more significant if analysed as a function of redshift, increasing up to a factor of $\simeq$30-50. Most importantly, as shown in the previous section, {\em JWST} data enable not only the detection of $z>10$ galaxies, but also an accurate estimate of their stellar mass.

\subsection{The mass--luminosity relation}

We revisit here the so-called mass-luminosity correlation for LBG and characterize for the first time whether it is valid, and  what is the scatter around the relation, by means of {\em JWST} data. 

Prior to {\em JWST}, numerous studies have relied on the UV stellar luminosities to estimate the stellar mass of high redshift galaxies, lacking rest-frame optical and near infrared data, by adopting an average (stellar) mass-(UV)luminosity correlation.
This correlation has been used to estimate the stellar mass function by converting the UV luminosity function \citep[e.g.][]{gonzalez11,song16,kikuchihara20}.

\begin{figure}[ht!]
    \centering
\includegraphics[width=9cm]{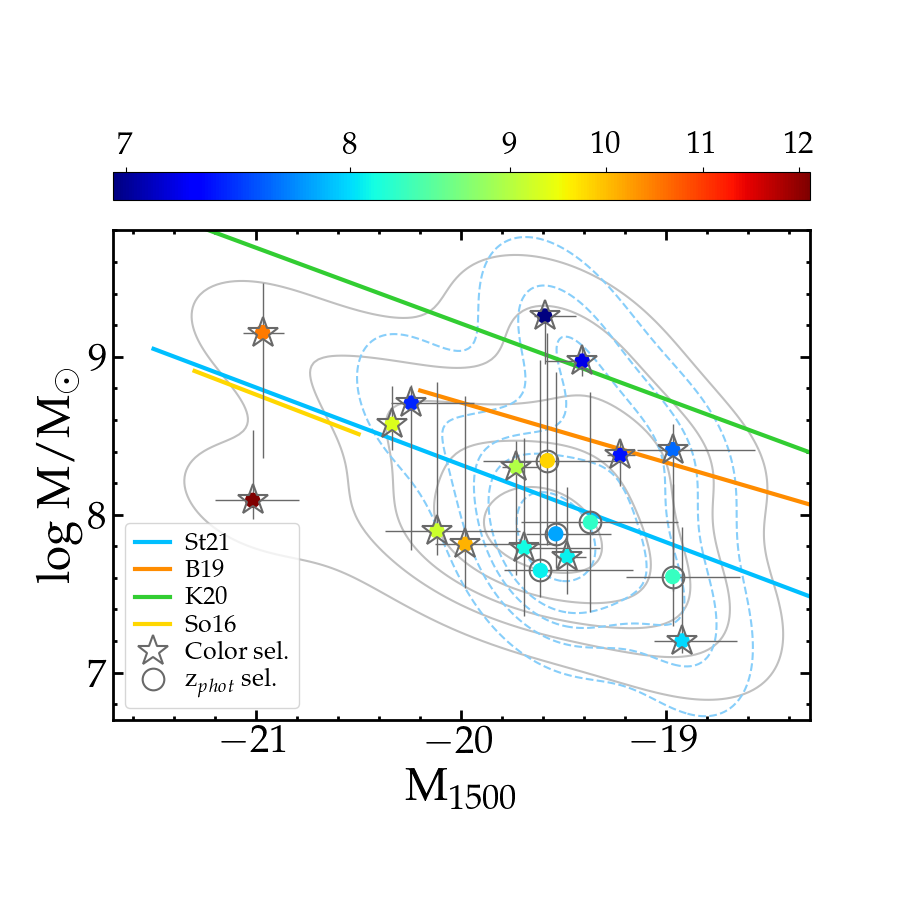}
\caption{Stellar mass as a function of observed absolute magnitude at 1500\AA, color-coded by redshift. Open stars and circles  show the candidates selected by means of colors and the additional ones added via photometric redshifts, respectively. 
Grey  solid curves and light blue dashed ones enclose 10, 30, 50, 70 and 90\%  probability densities of the total and of the $z<9$ sample, respectively. 
Colored  thick solid curves show mass-luminosity relations from the literature \citep{duncan14,song16,bhatawdekar19,kikuchihara20,stefanon21} 
at $z\sim8$, 
scaled to the same IMF and color-coded according to the legend, in the relevant range of luminosities.}
              \label{fig:massmuv}
\end{figure}

Fig.~\ref{fig:massmuv} shows the mass-luminosity relation compared with previous estimates from the literature.  At variance with previous works, our data do not show a correlation between mass and UV absolute magnitude, with a Pearson coefficient equal to -0.37 (a similar value is obtained limiting the fit to $z<9$ sources, for a cleaner comparison with previous studies).  
We note that the luminosity range probed by our data is limited due to the small volume of our observations, insufficient to include the  brighter and rarer sources. 
Nevertheless, 
the scatter at $M_{1500}>-20$ is very large (a factor of 50 in mass). 
This happens despite the fact that our sample is biased in favour of LBG and against galaxies with older/dustier stellar populations (none of our candidates has a best-fit E(B-V) larger than 0.2), whose search requires a different strategy.
Should these galaxies be detected in future {\em JWST} analysis, we expect the measured scatter in \ml~to further increase.

\subsection{The stellar mass-to-light ratio}

\begin{figure}[t!]
    \centering
\includegraphics[width=9cm]{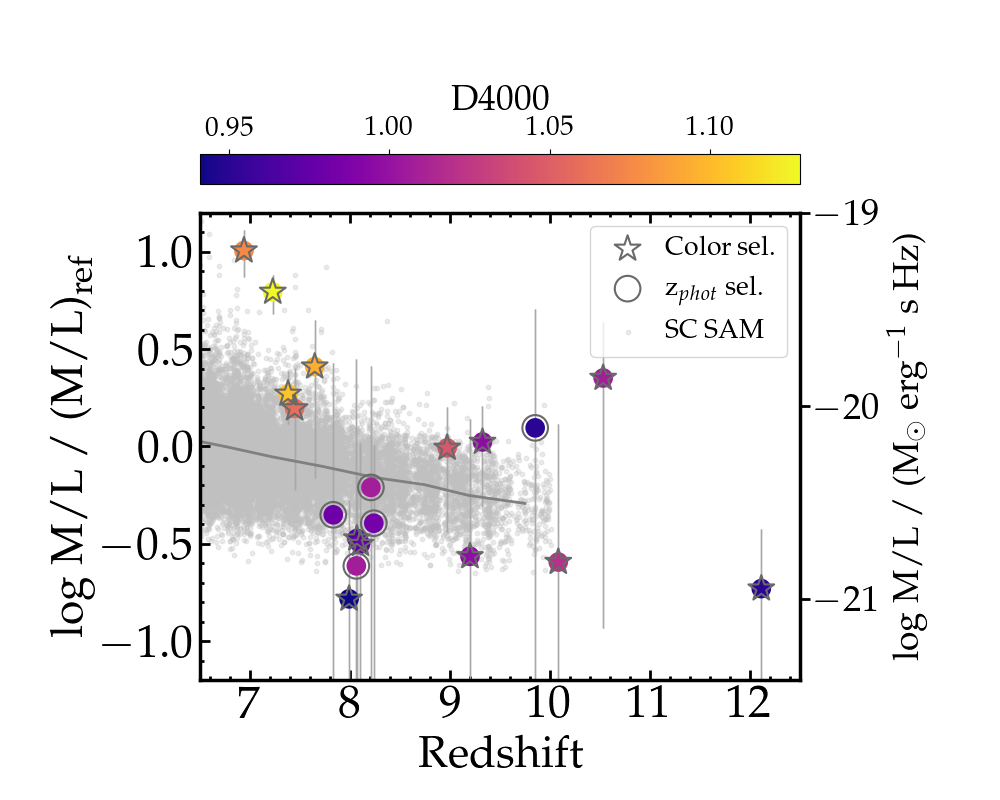}
\caption{Mass-to-light ratio at 1500\AA~as a function of redshift, color-coded by D4000. Open stars and circles show the color and the additional photo-$z$ selected candidates, respectively. The mass-to-light ratio is in units of the  \ml~of a 100 Myr old galaxy, with Solar metallicity, no dust, and constant SFH (see text). Small gray dots are the predictions from the Santa-Cruz semianalytic model \citep{somerville21}, with the dark gray curve showing the median predicted \ml~in bins of redshift.}
              \label{fig:ml}
\end{figure}

We show in Fig.~\ref{fig:ml} the mass-to-light ratio as a function of redshift. 
The mass-to-light ratio is plotted in units of a reference  \ml~(${M/L}_{\mathrm{ref}}$) calculated for a 100 Myr old galaxy, with Solar metallicity, no dust, and constant SFH (such a galaxy, according to \citealt{bc03} models, has a stellar mass of 7.9$\times 10^7$ M$_\odot$ and a 1500\AA~luminosity of 1.3$\times 10^{28}$~erg/s/Hz if normalized to a SFR of 1$M_\odot/yr$).

The mass-to-light significantly changes from galaxy to galaxy, spanning 
two orders of magnitude within our sample, consistent with what seen on Fig.~\ref{fig:massmuv}. This result implies that the high-$z$ galaxy population is largely heterogeneous, with galaxies observed in different evolutionary stages, as well as experiencing a broad variety of SFHs. 
In particular, the observed high \ml~suggest the presence of evolved stellar populations already existing at these high redshift, with tantalizing evidence that aggregation of baryons may proceed at a faster rate than predicted by galaxy formation models.

The  wide range of observed \ml~implies that previous results based on an average value for large galaxy samples should be revisited and revised, and might change significantly once {\em JWST} information is included. 

As expected, the mass-to-light ratio correlates 
with the amplitude of the D4000 break, 
encoded in the color map, with more  evolved systems showing higher \ml. Similar behavior is observed also with other galaxy properties that are indicative of the galaxy evolutionary stage, such as the specific SFR or age/2$\tau$\footnote{With the adopted parameterization, an age/2$\tau$ equal to 1 corresponds to the peak of the  SFH, with lower and higher values characterizing the increasing and  decreasing phase, respectively.}. 

Due to the large error bars, the limited size of the sample and incompleteness effects,  it is difficult to assess or exclude an evolutionary trend. 

To gain some insight on this issue, 
we plot in Fig.~\ref{fig:ml} the predictions from the Santa Cruz semianalytic model \citep{somerville21}, showing a decreasing trend all the way to $z=10$. Our data encompass and 
exceed the predicted scatter in \ml. The range of predicted \ml~in the model,  spanning a factor of $\sim$10, correlates with the specific SFR, likely due to the  combination of bursty and smooth SFHs. 
While observational uncertainties prevent us from drawing firm conclusions, the 
significantly larger scatter that we observe suggests that the simulations may underestimate the burstiness of mass growth.

\subsection{The observed cosmic Stellar Mass Density}\label{sec:smd}

We discuss here the accuracy of cosmic Stellar Mass Density (SMD) estimates obtained by converting luminosities through an average \ml~ratio. 

\begin{figure}[t!]
    \centering
\includegraphics[width=9cm,]{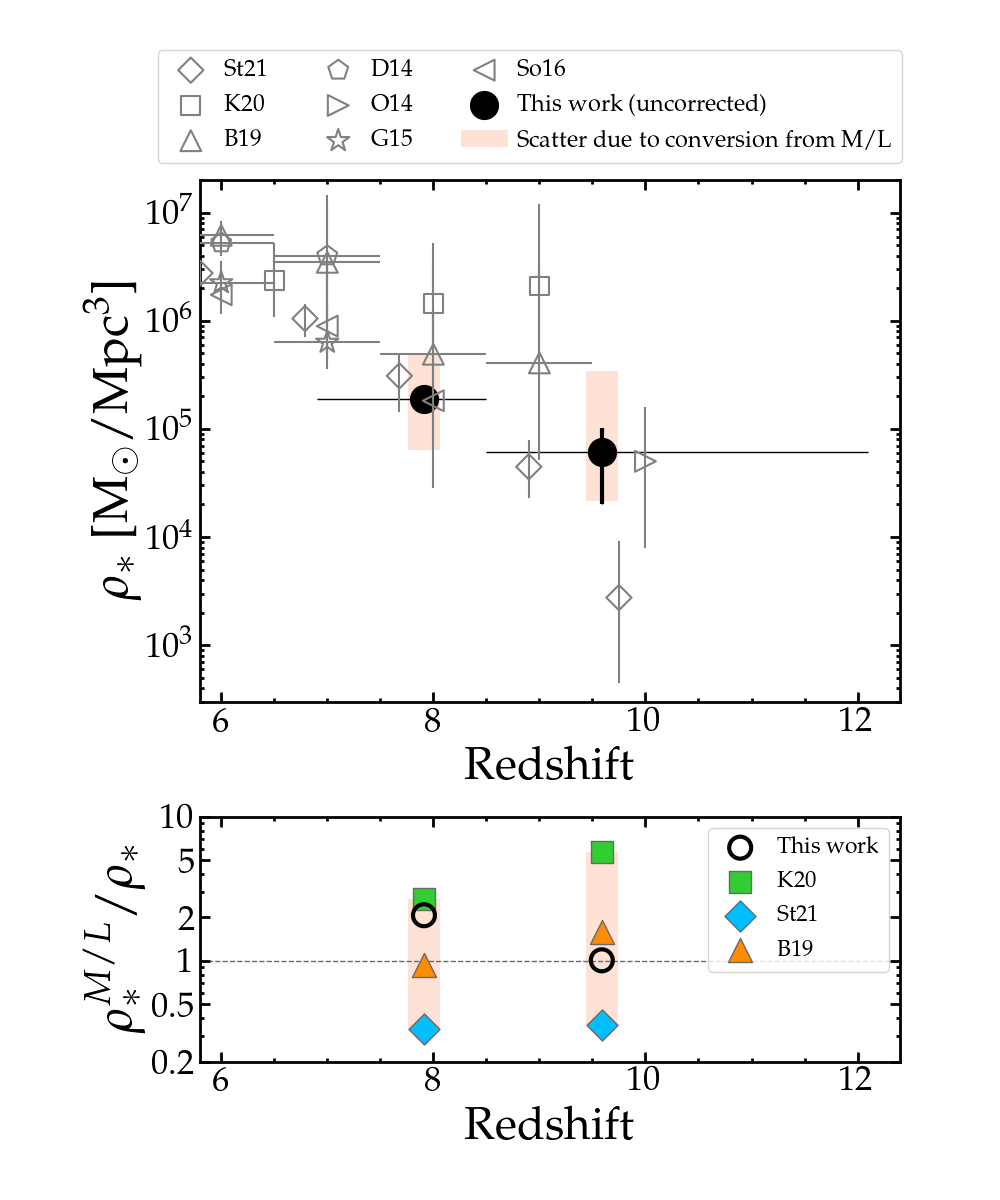}
\caption{{\it Upper} panel: Observed (i.e. uncorrected for incompleteness and lensing) cosmic Stellar Mass Density from this study (large solid black circles). Gray thin open symbols show a  collection of results from the literature (see legend; \citealt{duncan14}, \citealt{oesch14}, \citealt{grazian15}, \citealt{song16}, \citealt{bhatawdekar19}, \citealt{kikuchihara20}, \citealt{stefanon21}). The salmon shaded bar shows the range encompassed by indirect measurements of the SMD obtained by converting the 1500\AA~luminosities  
through several  mass-luminosity relations from the literature, as well as from the average value measured in this study. 
{\it Lower} panel: Ratio of the Stellar Mass Density indirectly inferred by converting 1500\AA~luminosities and the one directly obtained from  measured stellar masses. We considered the mass-luminosity relations  of \cite{kikuchihara20} (green squares), \cite{bhatawdekar19} (orange triangles),  \cite{stefanon21} (blue diamonds), and the average value of the \ml~calculated in each of the redshift bins from our data (black circles). 
}
\label{fig:md}
\end{figure}

We show in the upper panel Fig.~\ref{fig:md} a first attempt to infer the  SMD. We calculate the observed SMD ($\rho_*$) by adding up the stellar masses of all galaxies in two redshift bins ($6.9<z<8.5$ and $8.5<z<12.1$) and dividing by the cosmological volume of the bins.
The associated error bars are obtained by propagating the mass uncertainties.

Following a common approach at these redshifts, we also compute the SMD ($\rho_*^{M/L}$)  by converting UV luminosities through different mass-luminosity relations taken from the literature \citep{kikuchihara20,bhatawdekar19,stefanon21}, as well as by means of the mean \ml~measured in this study in each redshift bin. The salmon shaded regions on Fig.~\ref{fig:md} show the range of SMDs spanned by these indirect measurements. 

The ratio between these measurements and  $\rho_*$ is shown in the lower panel of Fig.~\ref{fig:md}. This ratio can be as high as a factor of $\sim$6, in both directions. The actual discrepancy depends on the mass and luminosity distributions of the specific data set, that may be different from the one on which the mass-luminosity relations have been estimated.  However,  we note that not even the average \ml~inferred on the same sample of galaxies is  always able to reproduce the measured SMD,  depending on the  distribution of sources within the redshift bin.

Fig.~\ref{fig:md} also reports  a collection of SMD measurements from the literature. 
These data show that, while a consensus has been reached at $z\lesssim 3$ \citep[e.g.][]{mcleod21}, the global picture is much more uncertain at redshifts above $z\sim 7$,  
where more than a factor of 10 variance exists among different studies. In this first paper, we do not attempt to compare our finding with the literature. We defer this comparison to future work once incompleteness and lensing magnification will have been properly characterized (these two effects move the points in opposite directions).
Nevertheless, we want to draw the attention to the level of uncertainty that can be introduced in the SMD when direct stellar mass measurements are not available. 
We have demonstrated that  the scatter among the results obtained with various estimates of the \ml~is a dominant source of uncertainty, that will require the full power of {\em JWST} to be contained.

\section{Summary and conclusions}\label{sec:summary}
 
This Letter presents a first analysis of the stellar masses and mass-to-light ratios of  galaxies above $z>7$. This analysis can only be performed thanks to the new GLASS-JWST NIRCam data that directly probe the  rest-frame optical flux of these galaxies. 
We show that with {\em JWST}  stellar masses of $z\gtrsim 7$ galaxy candidates can be measured with an accuracy that is  at least 5-10 
times better than previously possible, with considerably shorter exposure times. The observed UV \ml~spans   two orders of magnitude, 
revealing a broad variety of physical conditions in early galaxies. We demonstrate that previous assumptions of an average UV \ml, or mass-luminosity relation,  may introduce  
 systematic uncertainties in the cosmic SMD estimates that can be as high as a factor of  $\sim$6. 
This exploratory study demonstrates the power of {\em JWST} for studying high redshift galaxies and the assembly of their stellar mass. We will address this topic in more detail in a future analysis after completion of the GLASS-JWST program.

\begin{acknowledgments}
This work is based on observations made with the NASA/ESA/CSA James Webb Space Telescope. The data were obtained from the Mikulski Archive for Space Telescopes at the Space Telescope Science Institute, which is operated by the Association of Universities for Research in Astronomy, Inc., under NASA contract NAS 5-03127 for {\em JWST}. These observations are associated with program JWST-ERS-1324. The JWST data used in this paper can be found on MAST: http://dx.doi.org/10.17909/fqaq-p393. We acknowledge financial support from NASA through grant JWST-ERS-1324. KG and TN acknowledge support from Australian Research Council Laureate Fellowship FL180100060. MB acknowledges support from the Slovenian national research agency ARRS through grant N1-0238. PR acknowledges financial support through grants PRIN-MIUR 2017WSCC32 and 2020SKSTHZ.

\end{acknowledgments}

%

\vspace{5mm}








\bibliographystyle{aasjournal}



\end{document}